\documentclass[epj]{svjour}
\usepackage{url}
\usepackage{hyperref}
\usepackage{xspace}
\usepackage{graphicx}
\usepackage{amssymb}
\usepackage{multirow}
\usepackage{array}
\usepackage[usenames,dvipsnames]{color}
\usepackage{amsmath}
\usepackage[pagewise]{lineno}
\usepackage{color}
\usepackage{graphics}
\usepackage{cite}
\newcommand{\bef}{\begin{figure}}
\newcommand{\eef}{\end{figure}}
\newcommand{\bc}{\begin{center}}
\newcommand{\ec}{\end{center}}
\newcommand{\nuebar}{\ensuremath{\overline{\nu}_{e}}}
\newcommand{\antinue}{\ensuremath{\overline{\nu}_{e}}}

\journalname{Eur. Phys. J. C}

\begin{document}
\title{Sensitivity to sterile neutrino mixing using reactor antineutrinos}
\author{S. P. Behera\thanks {e-mail:shiba@barc.gov.in}, D. K. Mishra\thanks{dkmishra@barc.gov.in}, 
\and L. M. Pant
\institute{Nuclear Physics Division, Bhabha Atomic Research Centre, Mumbai 
- 400085, India}}
\date{Received: date / Accepted: date}

\abstract{
The reactor antineutrinos are used for the precise measurement of oscillation 
parameters in the 3-neutrino model, and also used to investigate active-sterile 
neutrino mixing sensitivity in the 3$+$1 neutrino framework. 
In the present work, we study the feasibility of sterile neutrino 
search with the Indian Scintillator Matrix for Reactor Anti-Neutrino (ISMRAN) 
experimental set-up using electron antineutrinos($\overline{\nu}_e$) produced from 
reactor as a source. The so-called 3$+$1 scenario is considered for active-sterile 
neutrino mixing, which leads to projected exclusion curves in the sterile neutrino 
mass and mixing angle plane. The analysis is performed considering both the reactor 
and detector related parameters. It is found that, the ISMRAN set-up can observe the 
active-sterile neutrino mixing sensitivity for $\sin^{2}2\theta_{14} \geq$ 0.064 and 
$\Delta m^{2}_{41}$ = 1.0 eV$^2$ at 90$\%$ confidence level for an exposure of 1 
ton-year by using neutrinos produced from the DHRUVA reactor with thermal power of 
100 MW$_{th}$. It is also observed that, there is a significant improvement of the 
active-sterile neutrino mixing parameter $\sin^{2}2\theta_{14}$ to $\sim$ 0.03 at the 
same $\Delta m^{2}_{41}$ by putting the ISMRAN detector 
set-up at a distance of 20 m from the compact proto-type fast breeder reactor (PFBR) 
facility with thermal power of 1250 MW$_{th}$.
}
\authorrunning{S. P. Behera et al.}
\maketitle
\section{INTRODUCTION}                                               %
\label{sec:intro}
The phenomena of neutrino oscillation have been established by several experiments 
using neutrinos from both natural (atmospheric and solar) and man-made (reactor and 
accelerator) sources. It shows the mixing between flavor and mass eigenstates, hence 
established that neutrinos have non-zero masses. Presently, the study of 
neutrino physics is in the precision era. However, experimental 
observations from various short baseline (SBL) experiments cannot be explained by 
3-neutrino mixing paradigm which requires new additional neutrino called as `sterile 
neutrino'. So the concept of this sterile neutrino could explain the results from 
GALLEX~\cite{GALLEX} and SAGE~\cite{SAGE} Gallium experiments, find a deficit in 
neutrino flux while calibrating the detectors with radioactive sources. They have 
reported that the ratio of numbers of observed to predicted events
is 0.88$\pm$0.05~\cite{Abdurashitov:2005tb} and it is known as ``Gallium anomaly". 
The accelerator based SBL experiments such as Liquid Scintillator Neutrino 
Detector (LSND)~\cite{Aguilar:2001ty} at a baseline of $\sim$ 30 m observed 
an unexplained excess of electron anti-neutrinos (\antinue) in a muon anti-neutrino
beam. The MiniBooNE experiment also observed similar excess in
$\overline{\nu}_{\mu} \rightarrow \overline{\nu}_{e}$ 
mode~\cite{Aguilar-Arevalo:2013pmq}. The recent MiniBooNE data
are consistent with the excess of events reported by the LSND.
The significance of the combined analysis of both the experiments
is an excess of 6.0$\sigma$~\cite{Aguilar-Arevalo:2018gpe}. There is an anomalous 
behavior has also been observed in the measurement of the reactor \antinue~ flux and 
spectrum. The precise energy spectrum of antineutrino flux produced by the reactors 
are recalculated by Mueller $et~al.$~\cite{Mueller:2011nm} which shows a 
significantly about 6$\%$ higher than experimental measurements at small distance.  
This discrepancy between the predicted and observed reactor antineutrino flux is 
known as the ``reactor antineutrino anomaly" (RAA)~\cite{Mention:2011rk}. There are
basically two possible explanations for this discrepancy. One is the incomplete
 reactor models or nuclear data due to underestimated systematics of the 
measurements of beta spectra emitted after
fission~\cite{Feilitzsch1982, Schreckenbach1985, Hahn1989} or of the conversion 
method~\cite{Mueller:2011nm,Huber:2011wv,Hayes2016,Huber2016}. The other explanation 
is an oscillation of \antinue~  into a fourth light sterile neutrino. 
Moreover, measurements of the reactor \antinue~ spectra show a discrepancy 
compared to predictions, particularly at energies of $\sim$ 5 MeV. The discrepancy in 
\antinue~ spectra is confirmed by  RENO~\cite{RENO:2015ksa}, Daya 
Bay\cite{An:2015nua}, Double Chooz~\cite{Abe:2015rcp}, and  NEOS ~\cite{Ko:2016owz} collaborations by measuring
the reactor \antinue~ energy spectrum. The distortion in energy spectra has been 
correlated to the reactor power~\cite{An:2015nua}, which may be due to the 
$^{235}$U fuel~\cite{An:2017osx}. In order to verify the existence of active to 
sterile neutrino oscillation hypothesis as the possible origin of the RAA and, also 
to clarify the origin of the bump at 5 MeV in the \antinue~ spectra,
there are several experiments underway and some will take data soon.

To address the RAA, the SBL experiments are aiming to measure the reactor \antinue~
spectrum at two or more different distances and trying to reconstruct
the \antinue~ survival probability both as a function of energy
and source to detector distance, $L$. 
independent of any reactor model prediction.
The $L$ dependence is what gives the cleanest signal in the case
of the sterile neutrino, and studying the ratio of the spectra
measured at two different distances allows to avoid 
the problem of the theoretical spectrum. The DANSS group has performed the experiment 
at 3 distances from reactor core varied from 10.7 m to 12 m to find out the 
active-sterile neutrino mixing by measuring the positron energy spectra. They have 
observed that the excluded area in the $sin^2 2\theta_{14} - \Delta m^{2}_{41} (= m_4^2 - m_1^2)$ plane 
covers a wide range of the sterile neutrino parameters up to $sin^2 2\theta_{14} <$ 
0.01~\cite{Alekseev:2018efk}. Similarly, the STEREO~\cite{Almazan:2018wln} group has 
measured the antineutrino energy spectrum in six different detector cells covering 
baselines between 9 and 11 meters from the compact core of the ILL research reactor.
Their results are compatible with the null oscillation hypothesis and the best fit 
of the reactor antineutrino anomaly is excluded at 97.5$\%$ confidence level.
Recently, PROSPECT group has measured the reactor \antinue~ spectra using a movable 
segmented detector array and their observation disfavors the RAA best fit point at 
2.2$\sigma$ C.L. and constrains significant portions of the previously allowed 
parameter space at 95$\%$ confidence level~\cite{Ashenfelter:2018iov}.

This paper presents the results of an investigation on finding a possible mixing of 
a single sterile neutrino with the 3 known active neutrinos, $viz.$ the (3~$+$~1) 
model. It is the only allowed active-sterile neutrino mixing 
scheme~\cite{Gariazzo:2017fdh} under the assumption of 4 neutrino model. At SBL, the presence of sterile neutrinos with 
squared mass difference $\Delta m_{41}^2 \sim$ 1 eV$^2$ leads to 
fast oscillations resulting the reduction of reactor $\antinue$ flux, otherwise 
absent in the standard 3-neutrino paradigm. This study quantifies 
the sensitivity of Indian Scintillator Matrix for Reactor Anti-Neutrino (ISMRAN) 
experimental set-up in constraining the active-sterile neutrino mixing parameters.  
In this work, we have considered various reactor ($viz.$ thermal power, core size, 
duty cycle, burn up) as well as detector response related parameters ($viz.$ energy 
resolution and detection efficiency) and also at several reactor core to detector 
distance while constraining active-sterile neutrino mixing at an exposure of 1 
ton-year.

The outline of the paper is as follows. A detailed description of the ISMRAN detector
set-up and the neutrino detection principle is discussed in Sec.~\ref{sec:detector} 
and in Sec.~\ref{sec:detectionprinciple}, respectively. The sterile neutrino 
oscillation formalism is introduced in Sec.~\ref{sec:osciprob}. The incorporation of 
detector resolutions on neutrino induced true events is discussed in 
Sec.~\ref{sec:simul}. The statistical analysis using both oscillated and without 
oscillated events based on $\chi^{2}$ estimation is given in Sec.~\ref{sec:chisq}. 
The sensitivity to sterile neutrino mixing at an exposure of 1 ton-yr is discussed in 
Sec.~\ref{sec:results}. Finally, in Sec.~\ref{sec:summary}, we summarize our findings 
and discuss the implication of this work. 

\section{ISMRAN DETECTOR}\label{sec:detector} %
The ISMRAN experimental set-up is being developed for detecting reactor 
$\overline{\nu}_e$, searching for possible existence of sterile neutrino and 
monitoring of reactor power at the DHRUVA reactor facility in Bhabha Atomic Research 
Centre (BARC), India. The ISMRAN detector set-up will consist of an array of 100 
plastic scintillator (PS) bars with weight of about 1 ton~\cite{Mulmule:2018efw}. The 
dimension of each PS bar is 100 
cm$\times$10 cm$\times$10 cm  wrapped with Gadolinium coated aluminized mylar foils. 
Each PS bar is coupled with two 3" Photo-multiplier tubes at both ends. The schematic 
of the detector set-up is shown in Fig.~\ref{fig:scintarray}. The advantage of the 
ISMRAN set-up is that, it is compact in size and maneuvered from one place to another 
easily. Also the segmented detector array can provide the additional position 
information while reconstructing the neutrino induced events and can improve the 
active sterile neutrino mixing sensitivity of the ISMRAN detector. To suppress both 
the natural and reactor related background, detectors are covered by a passive 
shielding material Lead (for gamma rays) of 10 cm thick and then followed by 10 cm 
thick of borated polyethylene (for neutrons). The detector is positioned at a 
distance of $\sim$13 m from the center of a cylindrical reactor core and can be moved 
closer to the core upto 7 m. The reactor has radius $\sim$1.5 m and height $\sim$3.03 
m (defined as an extended source)~\cite{Agarwal:dhruva}. The reactor can operate at 
a maximum thermal power of 100 MW$_{th}$ consuming natural uranium as fuel and 
producing about 10$^{19}$ $\antinue$/s. In future, it is planned to put the detector 
set-up at proto-type fast breeder reactor (PFBR) facility, IGCAR, Kalpakkam, 
India~\cite{Chetal:pfbr}. The PFBR has dimension of about 1 m  both in radius 
and height (defined as a compact source), and can operate at a maximum thermal power 
of 1250 MW$_{th}$. As the reactor is compact and produces higher thermal power, it is 
an ideal case to utilize the detector set-up for investigating the active-sterile 
neutrino mixing. With this experimental set-up, it can be possible to confirm or 
reject the existence of a light sterile neutrino by measuring the $\antinue$~ flux 
and energy spectra. At present a proto-type ISMRAN set-up of 1/5-th of the final 
detector volume which is under operation at DHRUVA reactor 
facility~\cite{Mulmule:2018efw}.
\bef[ht!]
\bc
\includegraphics[width=0.4\textwidth]{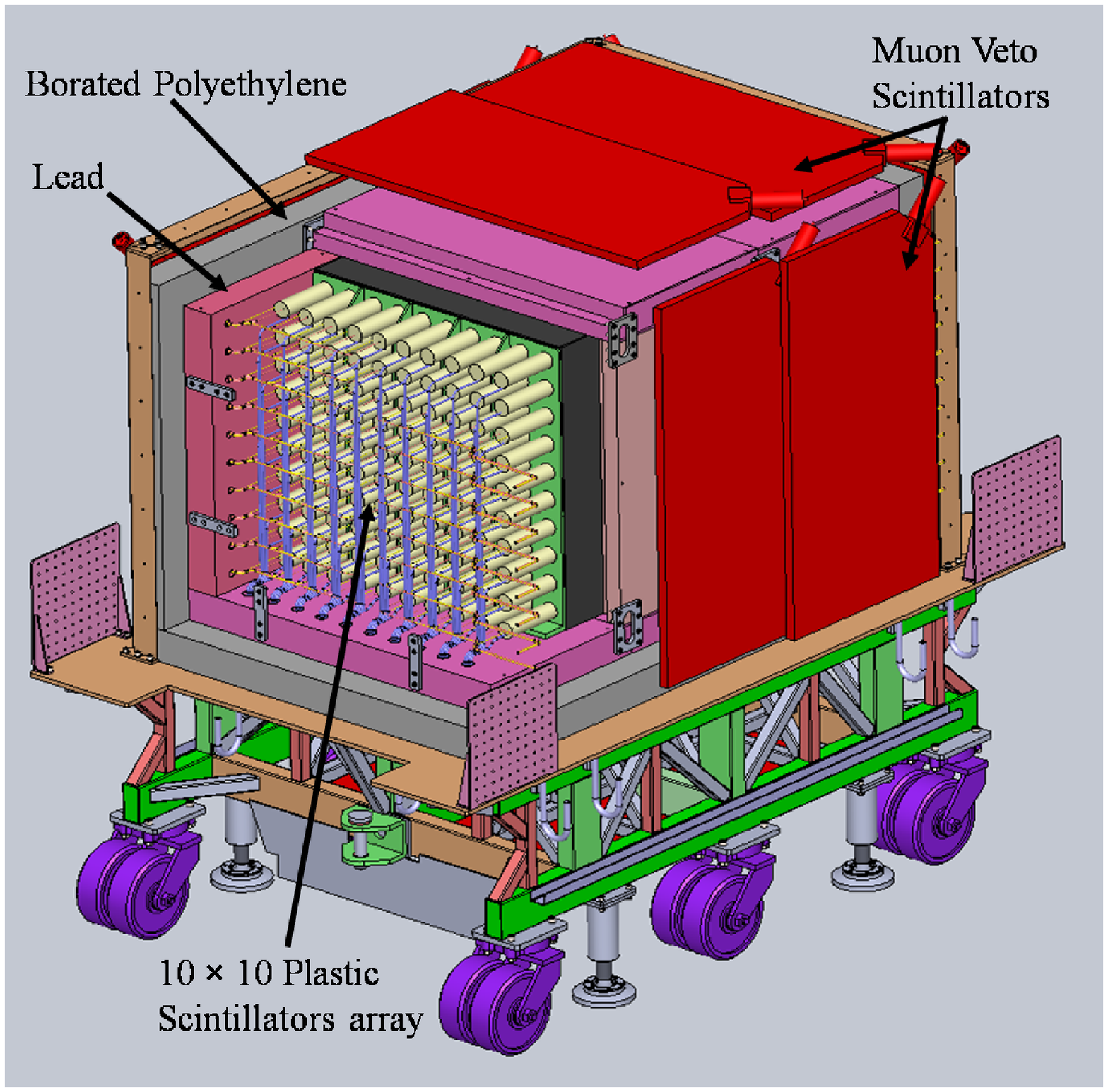}
\caption{Schematic of 100 cm $\times$ 10 cm $\times$ 10 cm plastic scintillators 
array with shielding materials for reactor antineutrino measurement.}
\label{fig:scintarray}
\ec
\eef
\section{\antinue~DETECTION PRINCIPLE}\label{sec:detectionprinciple}
The electron-antineutrinos produce from the reactor interact with protons in the PS 
bars, via the Inverse Beta Decay (IBD) process, 

\begin{equation}
\label{eq:ibdreac} 
\bar\nu_{e} + p \rightarrow n + e^{+}
\end{equation}
The Q-value of the above reaction is about $-$1.80 MeV and hence it limits the 
detection of antineutrinos. The positron which carries 
almost all of the available energy, loses it by ionization process in the detector 
and gets annihilated producing two gammas. The energy loss of the positron 
constitutes the `prompt’ signal along with the Compton scattered annihilated gammas 
given by 
 
\begin{equation}
\label{eq:prompt}
E_{prompt} = E_{\bar\nu_{e}} + Q + 2m_{e}c^{2},
\end{equation}
where E$_{\bar\nu_{e}}$ is the energy of electron-antineutrino. So from 
Eq.~\ref{eq:prompt}, it is observed that there is a one-to-one correspondence 
between the positron energy and $E_{\bar\nu_{e}}$. The neutron produced in 
Eq.~\ref{eq:ibdreac} carries a few keV's of energy and gets thermalized in collisions 
with protons in the PS bar. The neutron takes about 180$\mu s$ in order to gets 
captured  by proton in the PS bar produces gamma ray which is considered as a delayed 
signal. To further decrease the neutron captured time and improve the detector 
efficiency, PS bars are wrapped with Gadolinium coated aluminized mylar foil which 
has very high neutron captured cross-section. Further, the neutron captured time 
reduces to about 30--40 $\mu s$  and a cascade of gamma rays produce with total 
energy $\sim$8 MeV due to Gadolinium. The coincidence of a prompt positron 
signal and a  delayed signal from neutron captured by Gadolinium (Gd) uniquely 
identifies the IBD event.
\section{NEUTRINO OSCILLATION PROBABILITY WITH 3~$+$~1 MODEL}  %
\label{sec:osciprob}
The sterile neutrino oscillation probabilities are based on expansion of the 3 
generation Pontecorvo-Maki-Nakagawa-Sakata (PMNS)~\cite{Maki:1962mu} matrix to 
3$+$1 generation, where ``3" stands for three active neutrinos~($\nu_{e}$, 
$\nu_{\mu}$, $\nu_{\tau}$) and ``1" for a sterile neutrino~($\nu_{s}$). The neutrino 
flavors and mass eigenstates are related through

\begin{equation}
\begin{pmatrix}
\nu_{e}\\
\nu_{\mu}\\
\nu_{\tau}\\
\nu_{s}
\end{pmatrix}
=
\begin{pmatrix}
U_{e1} & U_{e2} & U_{e3} & U_{e4} \\
U_{\mu 1} & U_{\mu 2} & U_{\mu 3} & U_{\mu 4} \\
U_{\tau 1} & U_{\tau 2} & U_{\tau 3} & U_{\tau 4} \\
U_{s1} & U_{s2} & U_{s3} & U_{s4} \\
\end{pmatrix}
\begin{pmatrix}
\nu_{1}\\
\nu_{2}\\
\nu_{3}\\
\nu_{4}
\end{pmatrix}
\label{eq:mixingmatrix}\textrm,
\end{equation}
where $U$ is a unitary mixing matrix. In this analysis the following parametrization 
for $U$ has been considered

\begin{equation}
\label{eq:rotmatrix}
U = R(\theta_{34})R(\theta_{24})R(\theta_{23})R(\theta_{14})R(\theta_{13})
R(\theta_{12}),
\end{equation} 
where $R(\theta_{ij})$ are the (complex) rotation matrices, $\theta_{ij}$ are 
the mixing angles with $i,j$ = 1, 2, 3, 4; and the order of rotation angles 
are considered as given in Ref.~\cite{Palazzo:2013bsa}. Using the above definition, 
the flavor change can be described as a function of the mixing matrix elements and 
masses in terms of the neutrino oscillation probability

\begin{equation}
\begin{split}
P_{\alpha\beta} = \ & \delta_{\alpha\beta}-4\sum_{i>j}Re(U_{\alpha i}
U^{*}_{\beta i}U^{*}_{\alpha j}U_{\beta j})\sin^{2}\bigg(\frac{\Delta 
m^{2}_{ij}L}{4E_{\nu}}\bigg)\\ & + 2 \sum_{i>j}Im(U_{\alpha i}U^{*}_{\beta 
i}U^{*}_{\alpha j}U_{\beta j})\sin^{2}\bigg(\frac{\Delta m^{2}_{ij}L}{2E_{\nu}}\bigg),
\end{split}
\label{eq:prob}
\end{equation}
where $\alpha$, $\beta$ correspond to $e$, $\mu$, $\tau$, s; $\Delta m^{2}_{ij}$ = 
$m^{2}_{i}- m^{2}_{j}$ with $i>j$, $L$ is the source to detector distance in 
`meter' and $E_{\nu}$ is the energy of neutrinos in `MeV'. The oscillation 
probabilities for antineutrinos can be obtained by replacing mixing matrix elements 
$U$s with its complex conjugate ($U^{\ast}$s). Since 
Eq.~\ref{eq:rotmatrix} is independent of the CP-violating phases as they are not 
observable at SBL reactor setups, the third term in Eq.~\ref{eq:prob} will be 
zero~\cite{Palazzo:2013bsa}. For a small value of mixing angle $\theta_{14}$ and 
source to detector distance of few meters ($<$ 100~m), the oscillation from 
3$\times$3 mixing parameters can be neglected. Hence, the electron antineutrino 
survival probability in Eq.~\ref{eq:prob} is approximated to  

\begin{equation}
\label{eq:Detected2}
P_{\nu_e\nu_e}(E,L) \simeq 1-\sin^22\theta_{14}\sin^2\left(\frac{1.27 
\Delta m^2_{41}{L}}{E_{\nu}}\right),
\end{equation}
The analysis of 3$+$1 generation is reduced to that of two flavor framework with the 
oscillation parameters $\Delta m^2_{41}$ and $\sin^22\theta_{14}$ are given 
by
\begin{equation}
\Delta m^2_{41}  =  m^{2}_{4}- m^{2}_{1}~;~~
\sin^22\theta_{14}  =  4 |U_{e4}|^2(1 - |U_{e4}|^2), 
\end{equation}
where $U_{e4}$ = $\sin\theta_{14}$.
%
\section{SIMULATION PROCEDURE}
\label{sec:simul}     
\begin{table*}[ht]
 \begin{center}
\caption{\label{tab:para}{Fractional contribution of each element 
to reactor power and parameters used to fit the neutrino spectrum}}
\begin{tabular}{ cccccccc}
  \hline
  Element & $a$ & $b_0$ & $b_1$ & $b_2$ & $b_3$ & $b_4$& $b_5$\\
\hline
 $^{235}$U & 0.58 & 4.367 & -4.577 & 2.1 & -0.5294 & 0.06186 & -0.002777 \\

 $^{239}$Pu & 0.30 & 4.757 & -5.392 & 2.63 & -0.6596 & 0.0782 & -0.003536 \\ 

 $^{241}$Pu & 0.05 & 2.99 & -2.882 & 1.278 & -0.3343 & 0.03905 & -0.001754 \\

 $^{238}$U & 0.07 & 4.833 & 1.927 & -1.283 & -6.762 & 2.233 & -1.536 \\

  \hline
\end{tabular}
\end{center}
\end{table*}
The active-sterile neutrino mixing sensitivity of ISMRAN set-up will
be explored at DHRUVA as well as PFBR reactor facilities. The number of
neutrinos produced from the reactor depends on the thermal power. It is essential 
to know the fuel compositions contributing to the thermal power of the reactor. In 
order to estimate the number of $\antinue$ induced events produced in the detector, 
assumed parametrization for antineutrino flux considered in the analysis is as 
follows
\begin{equation}
f(E_{\antinue}) = \sum_{i =~0}^{4} a_{i} \exp\bigg(\sum_{j =~0}^{6} b_{j} 
E_{\antinue}^{j-1}\bigg),
\end{equation}
where `$a_i$' is the fractional contribution from $i$th isotope to the reactor 
thermal power,`$b_j$'s are the constant term used to fit the neutrino spectrum and 
$E_{\antinue}$ is neutrino energy in MeV. For DHRUVA reactor, we have assumed that 
the fractional contribution for each isotope to the reactor thermal power 
as given in Ref.\cite{Zhan:2008id} and the list 
of parameters used to fit the $\antinue$ spectrum due to $^{235}$U, $^{239}$Pu and 
$^{241}$Pu are considered from Ref.~\cite{Huber:2011wv} and for $^{238}$U is taken 
from Ref.~\cite{Mueller:2011nm}. The list of parameters used in this analysis are 
 listed in Table~\ref{tab:para}. Similarly, for PFBR we have considered the 
 fractional contributions due to  $^{235}$U and $^{239}$Pu are 70\% and 30\%, 
respectively~\cite{Schneider}. We 
have also considered the neutrino flux variation due to a finite size cylindrical 
reactor which depends on its radius and height as follows~\cite{neu_flux},

\begin{equation}
\phi = \phi_{0}~J_{0}(2.405 r/R)~cos(\pi z/H)
\end{equation}   
where $\phi_0$ is flux at the center of the reactor core, $R$ and $H$ are the 
physical radius and height of the cylinder, respectively, $J_0$ is the zeroth order 
Bessel function of first kind with $r$ ($0\leq r\leq R$) and z ($ 0 \leq z \leq H$) 
are the vertex position of the produced neutrinos in the reactor.
The leading order interaction cross-section~\cite{Vogel:1999zy} of $\nuebar$ for the 
IBD process is given by

\begin{equation}
\sigma_{IBD} = 0.0952 \times 10^{-42} \mathrm{cm}^2 (E_{e^{+}}~ p_{e^{+}} /1\mathrm{MeV}^2),
\end{equation}
where $E_{e^{+}}$ = $E_{\nuebar} - (m_n -m_p)$ is the positron energy with 
neglecting recoil neutron energy and 
$p_{e^{+}}$ is the positron momentum. It can be mentioned here that the neutrino is a 
neutral particle and can not be detected directly. In the detector we measure the 
neutrino induced charged particle, for the present case it is $e^+$. The 
detector resolution on true positron energy (kinetic) spectrum is incorporated 
assuming a standard Gaussian form of the energy resolution:

\begin{equation}
R(E,E_T) = 
\frac{1}{{\sqrt{2\pi}}\sigma} \exp(-\frac{(E - E_T)^2}{2\sigma^2})\,.
\label{Ereso}
\end{equation}
\noindent Here $E_T$ and $E$ are true and the measured positron energy, respectively.
The detector resolution considered in this study is in the form of $\sigma/E ~\sim$ 
20$\%$/$\sqrt{E}$. In the analysis, the neutrino induced events are distributed in 
terms of positron energy spectrum. We have considered 
total 80 bins in the $e^+$ energy range of 0--8 MeV. The number of events in $i$-th 
energy bin after incorporating the detector resolution is given as

\begin{equation}
N_{i}^{r} = \sum_{k} K_{i}^{k}(E_{T}^{k}) n_{k}
\end{equation}
The index $i$ corresponds to the measured energy bin and $N_{i}^{r}$ corresponds to 
the number of reconstructed events, $k$ is summed over the true energy of positron 
and $n_{k}$ is the number of events in $k$-th true energy bin. Further, $K_{i}^{k}$ 
being the integral of the detector resolution function over the $E$ bins and is given 
by

\begin{equation}
K_{i}^{k} = \int_{E_{L_{i}}}^{E_{H_{i}}} dE \frac{1}{\sqrt{2\pi\sigma_{E}^{2}}} 
e^{-{\frac{\left(E_{T}^{k}-E\right)^{2}}{2\sigma_{E}^{2}}}}
\end{equation}
The integrations are performed between the lower and upper boundaries of the measured 
energy ($E_{L_i}$ and $E_{H_i}$) bins. After incorporating detector energy 
resolution on neutrino induced events, both unoscillated and oscillated event 
distributions as a function of energy are shown in Fig.~\ref{fig:eventdist} assuming 
25$\%$ of the detection efficiency, 80$\%$ fiducial volume of the detector, 70$\%$ 
reactor duty cycle and, for an exposure of 1 ton-year while placing ISMRAN set-up at 
a distance of 13 m from the reactor core. The neutrino oscillation probability from 
one flavor to another not only depends on precise measurement of the source to 
detector distance but also on energy of neutrinos. The uncertainty in distance 
traveled by neutrino should be less than the oscillation wavelength in order to 
avoid a washout of the oscillation signal. Figure~\ref{fig:eventratio} shows the 
comparison of oscillated to unoscillated event ratios with and without incorporating 
the detector resolution as well as varying the source to detector distance as a 
function of $L/E_{\nu}$. Oscillated events are estimated by considering best fit 
values of active-sterile neutrino mixing parameter at $\sin^22\theta_{14}$ = 0.062 
and $\Delta m^2_{41}$ = 1.7 eV$^2$~\cite{Gariazzo:2017fdh}. In 
Fig.~\ref{fig:eventratio}, the black dotted 
line shows the event ratios in an ideal case $i.e.$ without incorporating the 
detector response and at a fixed source to detector distance. The red solid line 
shows the event ratios by considering a cylindrical reactor core where the position 
of the reactor core is generated using a Monte-Carlo method and point detector. It 
is observed that with the variation of source to detector distance, the neutrino 
oscillation probability washed out as compared to fixed path length. 
Also by varying the path length due to random vertices of reactor core and 
incorporating the detector resolution on neutrino energy spectrum, the oscillation 
probability has further washed out as shown by blue dashed line. It is to be 
mentioned here that rest of the studies are performed using the 
randomized vertex in the reactor core and 80\% fiducial volume of the detector.
\bef[ht!]
\bc
\includegraphics[width=0.4\textwidth]{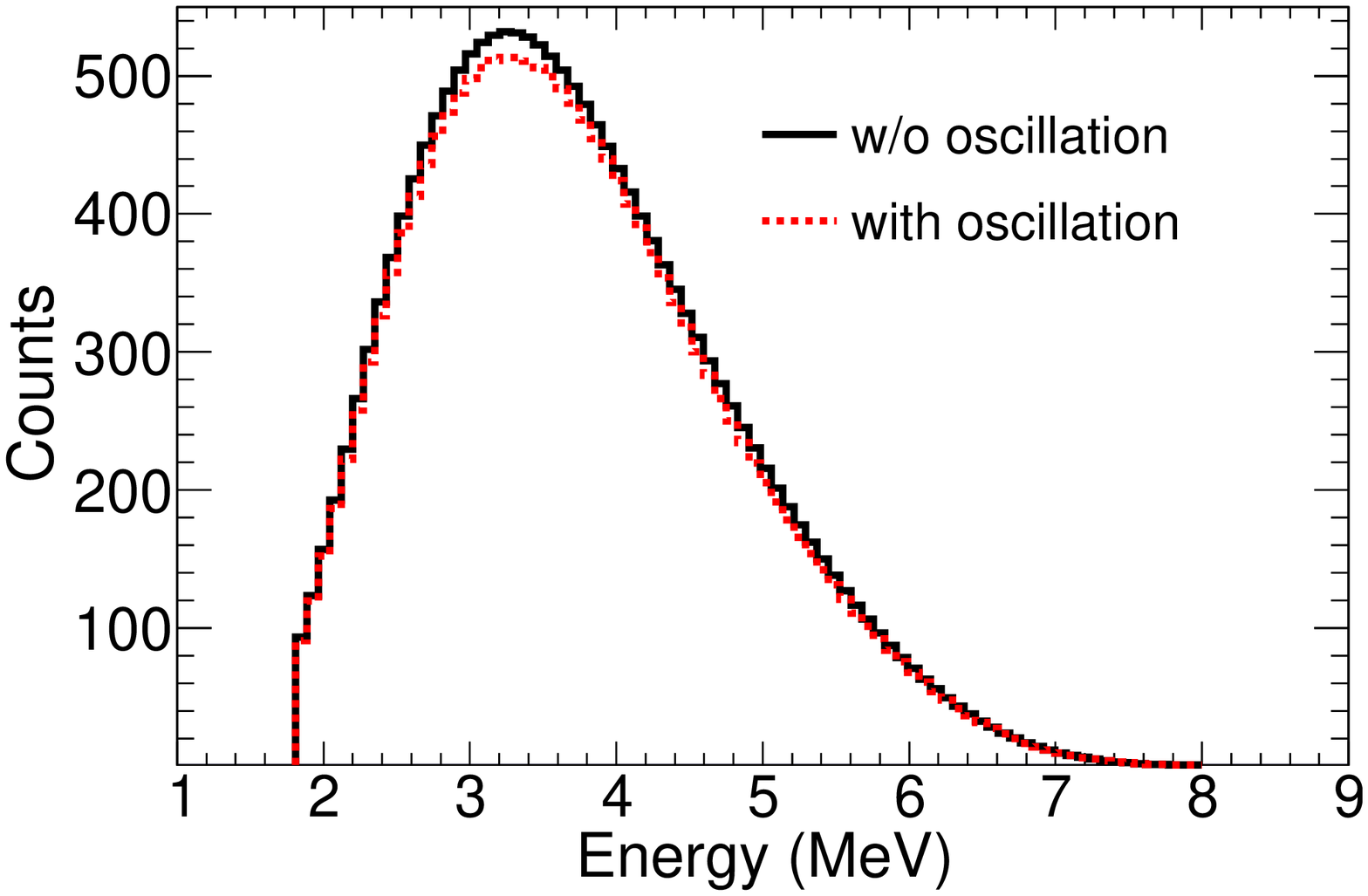}
\caption{Simulated event distribution without and with active-sterile neutrino 
oscillation after incorporating detector response. Oscillated events are estimated 
with  $\Delta m_{41}^2$ = 1.7 eV$^2$ and $\sin^22\theta_{14}$ = 0.062.}
\label{fig:eventdist}
\ec
\eef

\bef[ht!]
\bc
\includegraphics[width=0.4\textwidth]{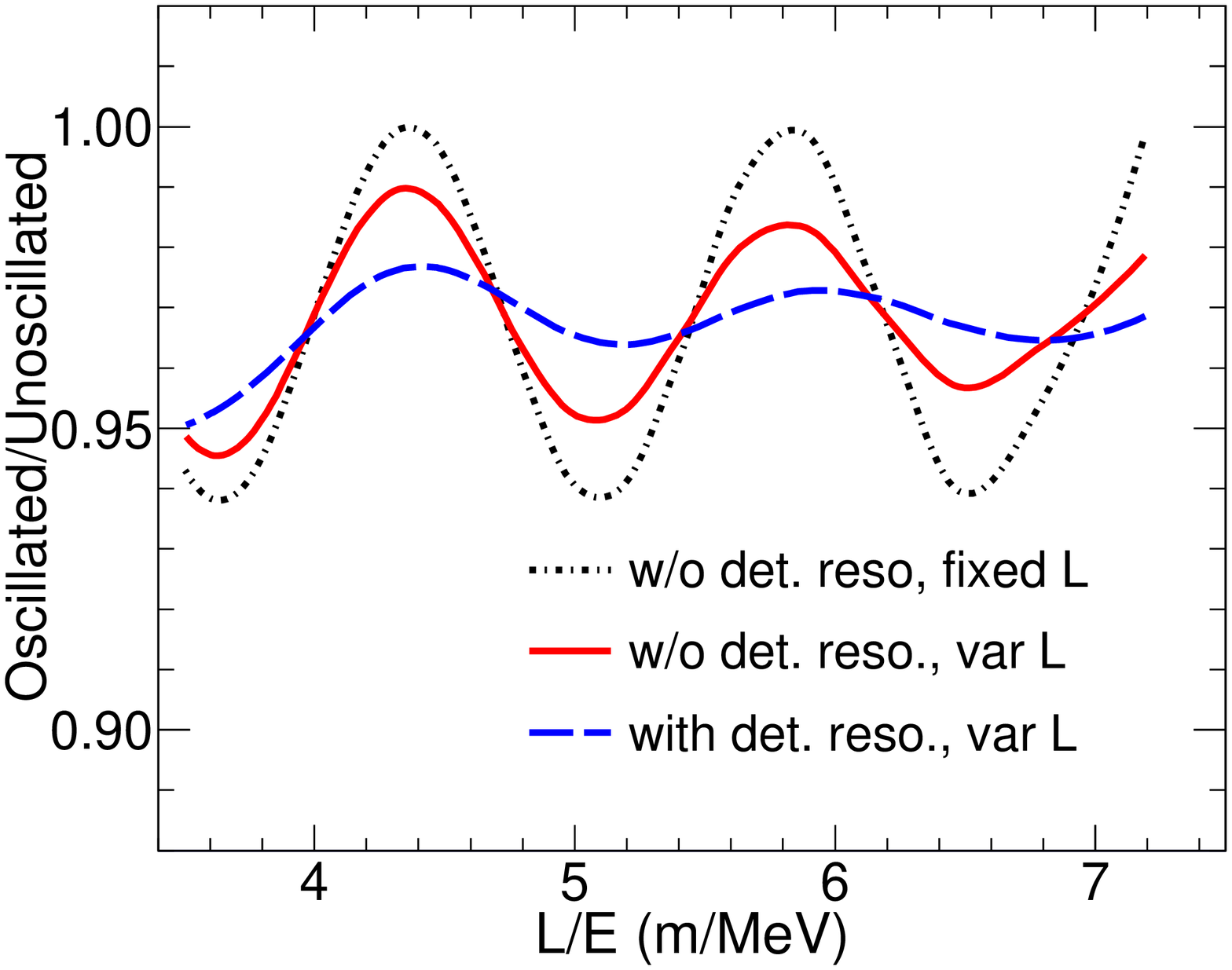}
\caption{Ratios of oscillated to unoscillated event distribution as a function $L/E$. 
Oscillated events are estimated with $\Delta m_{41}^2$ = 1.7 eV$^2$ and $\sin^2 
2\theta_{14}$ = 0.062. The black dotted line shows the ratios in ideal case $i.e.$ 
without incorporating the detector response and at a fixed path length of 13 m. The 
red solid line shows the events ratios by considering a cylindrical reactor core 
(parameters mentioned earlier) without incorporating the detector resolution and the 
same with detector resolution is shown in blue dashed line.}
\label{fig:eventratio}
\ec
\eef
\section{SENSITIVITY ESTIMATION OF SIMULATED DATA}
\label{sec:chisq}          %
In order to quantify the sensitivity of ISMRAN experimental set-up to the 
active-sterile neutrino mixing parameters $\theta_{14}$ and $\Delta m^2_{41}$, we 
perform the 
statistical analysis of event distribution for an exposure of 1 ton-year. After 
incorporating the detector response on the number of events estimated by 
considering with and without active-sterile neutrino oscillation, the sensitivity to 
the sterile neutrino mixing parameters has been obtained by calculating the 
$\chi^{2}$. To determine the exclusion limit for a given confidence interval at each 
value of $\Delta m_{41}^2$ we have scanned over the values of $\sin^2 
2\theta_{14}$ to simulate active-sterile neutrino oscillated event spectrum, and 
determine the boundary of the corresponding $\chi^2$ (e.g. $\chi^2$ = 4.61 for 90\% 
confidence limit(C.L.)). The $\chi^{2}$ can be defined as follows~\cite{pu}

\begin{equation}
\chi^{2} =\sum_{n=0}^{N} \bigg(\frac{ 
R_{n}^{th}-R_{n}^{ex}}{\sigma(R_{n}^{ex})}\bigg)^{2}
\label{eq:chi1}
\end{equation}%
 where $n$ is the number of energy bins, $R_{n}^{ex}$, $R_{n}^{th}$ are with 
oscillated and without oscillated (or theoretically predicted) events, respectively. 
The $R_{n}^{th}$ carries the information about systematic uncertainties  given by

\begin{equation}
R_{n}^{th}=R_{n}^{'th}\bigg(1+\sum_{i=0}^{k}\pi_{n}^{i}\xi_{i}
\bigg)+\mathcal{O}(\xi^{2})
\label{eq:chi2}
\end{equation}
with $\pi_{n}^{i}$ being the strength of the coupling between the pull variable 
$\xi_{i}$ and $R_{n}^{'th}$. Equation~(\ref{eq:chi1}) is minimized with respect to 
pull variables. Four systematic uncertainties such as 3$\%$ normalization 
uncertainty (including reactor total neutrino flux, number of target protons, and 
detector efficiency), nonlinear energy response of the detector by 1$\%$, uncertainty 
in energy calibration by 0.5$\%$. We have also considered the possibility of an 
uncorrelated experimental bin-to-bin systematic error of 2$\%$ which could result 
from insufficient knowledge of some source of background~\cite{Huber:2003pm}.
\bef[ht!]
\bc
\includegraphics[width=0.5\textwidth]{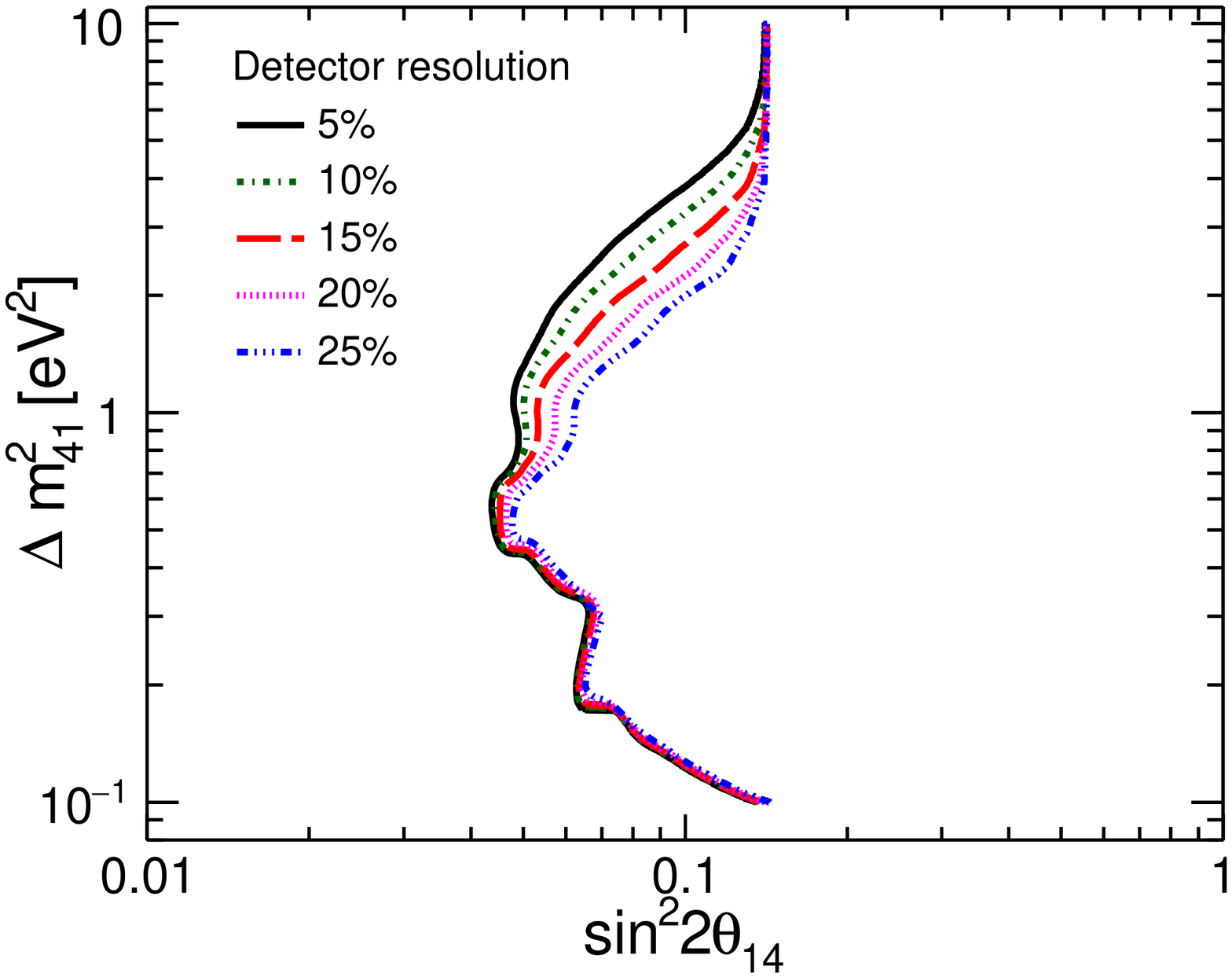}
\caption{The 90\% C.L. exclusion limits in the $\Delta 
m_{41}^2-\sin^22\theta_{14}$ plane, where $\sin^{2}2\theta_{14}= 4U^{2}_{e 
4}(1-U_{e4})^{2}$, expected from 1 ton-yr of the data at different PS detector 
resolution.}
\label{fig:diffreso}
\ec
\eef
\bef[ht!]
\bc
\includegraphics[width=0.5\textwidth]{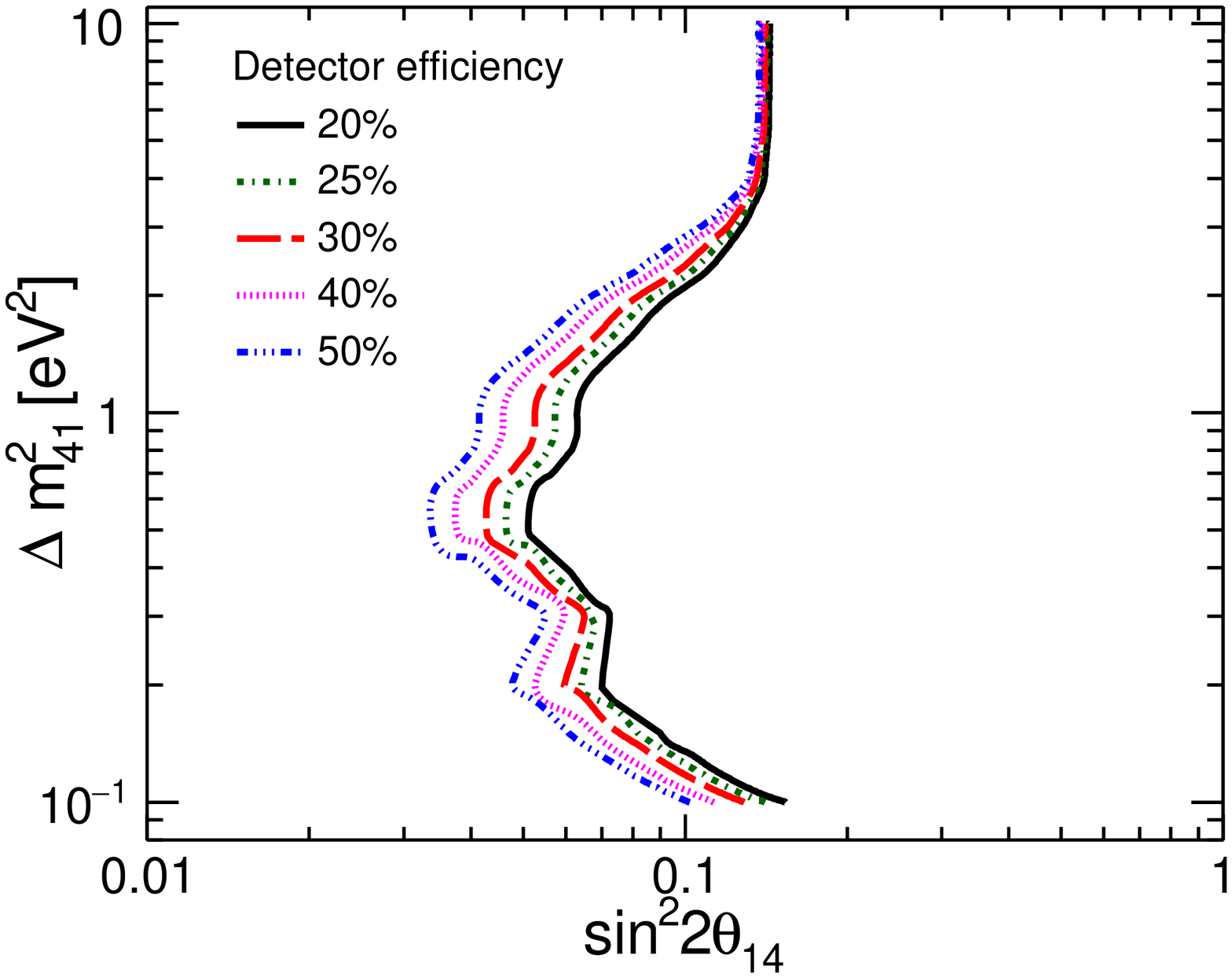}
\caption{The 90\% C.L. exclusion limits in the $\Delta 
m_{41}^2-\sin^22\theta_{14}$ plane, where $\sin^{2}2\theta_{14}= 4U^{2}_{e
4}(1-U_{e4})^{2}$, expected from 1 ton-yr of the data at various detector 
efficiencies.}
\label{fig:deteffi}
\ec
\eef

\section{RESULTS AND DISCUSSIONS}
\label{sec:results}  
The active-sterile neutrino mixing sensitivity depends on various reactor and 
detector parameters. As mentioned earlier the reactor related parameters are such as 
the thermal power, its fuel components, duty cycle and the core size. Apart 
from the reactor parameters, active-sterile neutrino mixing sensitivity also depends 
on detector mass, its fiducial volume, energy resolution, and detection efficiency. 
The simulation has been carried out by varying above mentioned parameters while 
finding the active-sterile neutrino mixing sensitivity as discussed below.

\bef[ht!]
\bc
\includegraphics[width=0.5\textwidth]{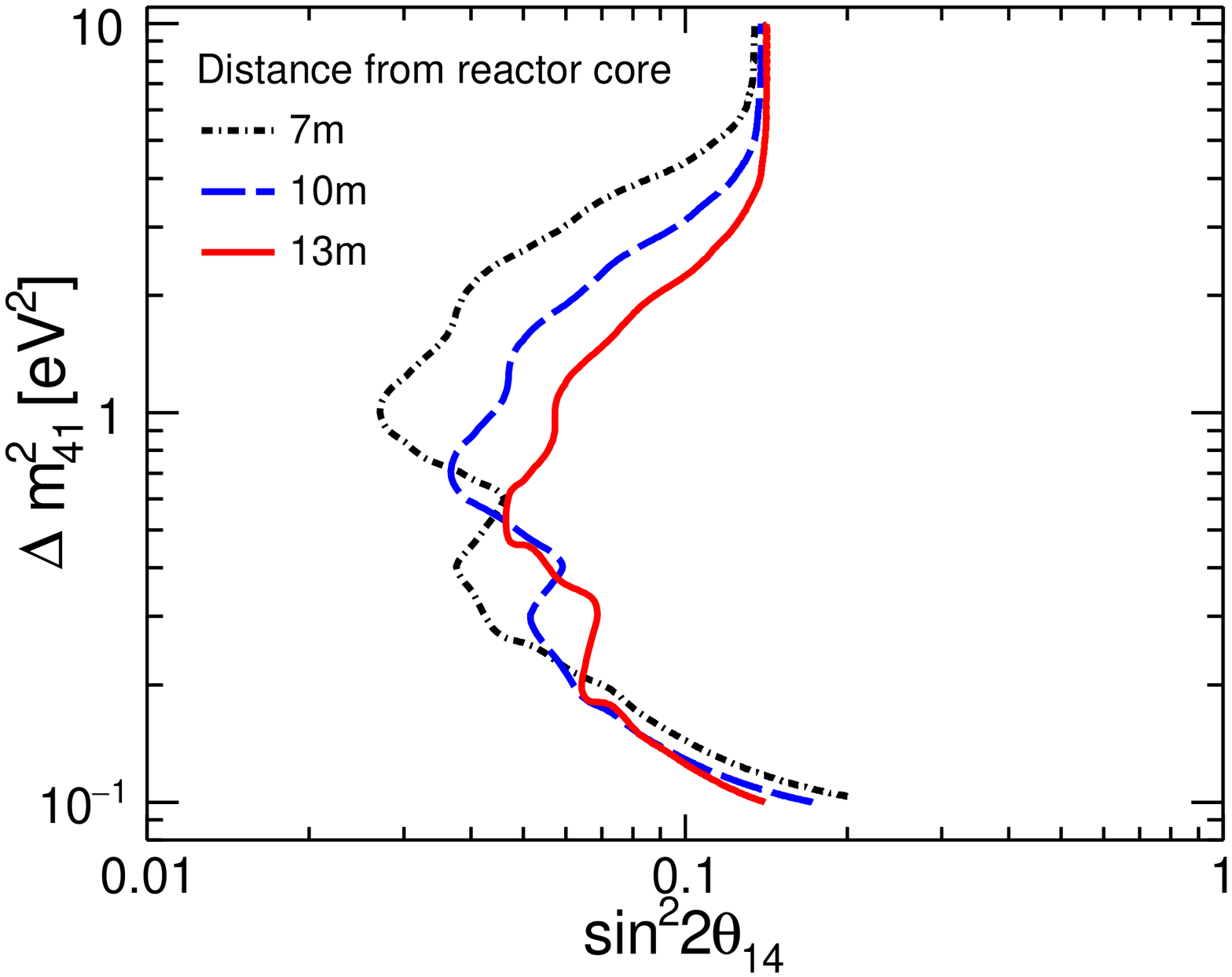}
\caption{The 90\% C.L. exclusion limits in the $\Delta m_{41}^2-\sin^22\theta_{14}$ 
plane, where $\sin^{2}2\theta_{14}= 4U^{2}_{e 4}(1-U_{e4})^{2}$, expected from 1 
ton-yr of the data at different source to detector path lengths with 100 MW$_{th}$ 
reactor power.}
\label{fig:diffpaths}
\ec
\eef

\bef[ht!]
\bc
\includegraphics[width=0.5\textwidth]{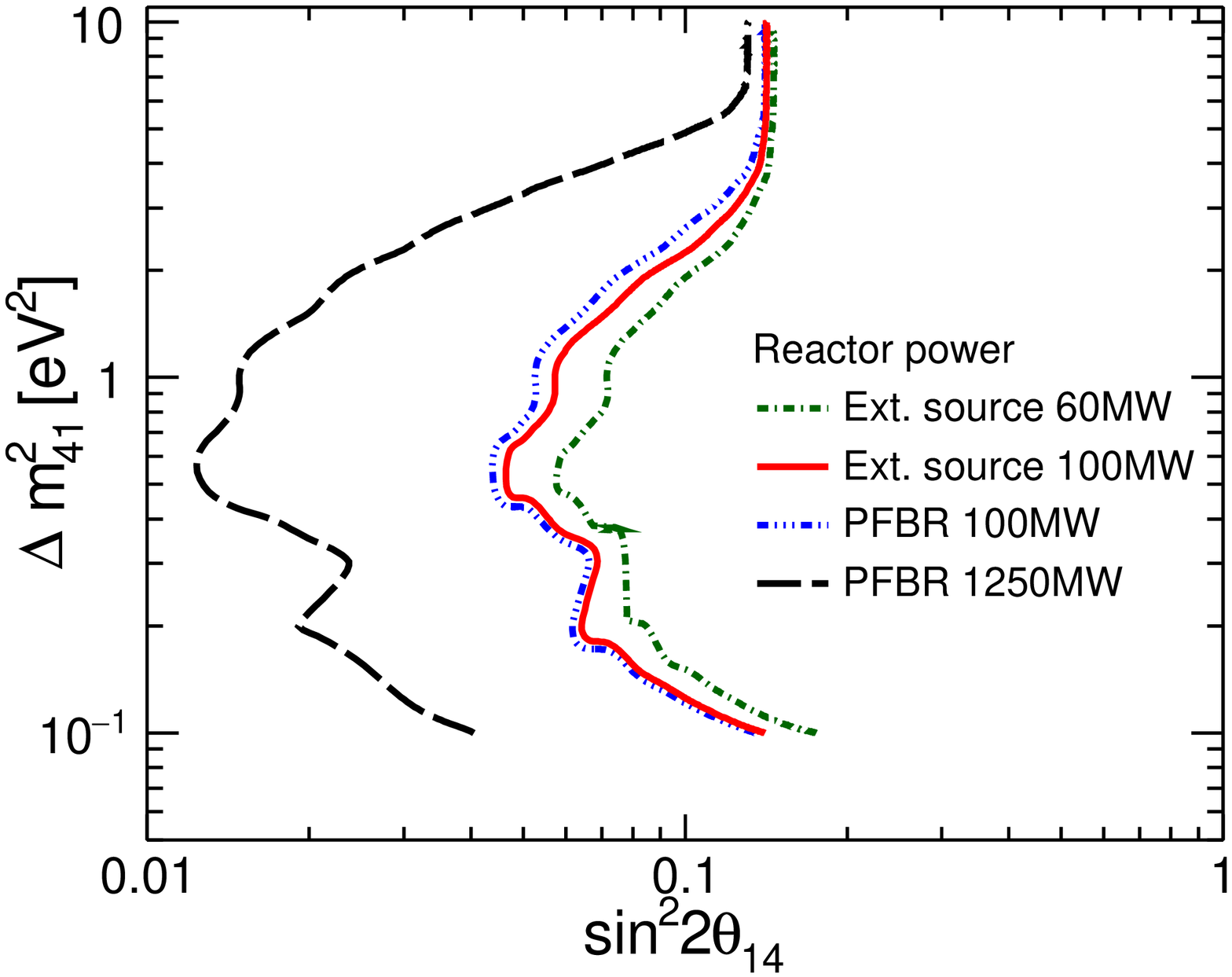}
\caption{The 90\% C.L. exclusion limits in the $\Delta m_{41}^2-\sin^22\theta_{14}$ 
plane, where $\sin^{2}2\theta_{14}= 4U^{2}_{e 4}(1-U_{e4})^{2}$, expected from 1 
ton-yr of the data at different reactor thermal power.}
\label{fig:diffpower}
\ec
\eef
\subsection{DETECTOR RESPONSE}
The oscillation probabilities of \antinue depend on the active-sterile 
neutrino mixing parameters such as angle and squared mass difference. The 
sensitivity of both these parameters depends on detector response such as resolution 
and efficiency. The upper limit for the active-sterile neutrino mixing angle 
$\theta_{14}$ for an exposure of 1~ton-yr is shown in Fig.~\ref{fig:diffreso} at 
90$\%$ C.L. in the $\Delta m_{41}^2$ - $\sin^22\theta_{14}$ plane considering 
different detector resolutions, $\sigma/E$ = 5\%--25$\%$/$\sqrt{E}$. The analysis is 
carried out considering reactor thermal power of 100 MW$_{th}$ produced from the 
extended reactor core and at 70$\%$ of its duty cycle. The detector is placed at a 
distance of about 13 m from the center of the reactor core. The detector has better 
active-sterile neutrino mixing sensitivity for resolution of $\sigma/E$ = 
5\%$/\sqrt{E}$. It is observed that at $\Delta m_{41}^2 < 0.5$ eV$^2$, 
active-sterile neutrino mixing sensitivity is independent of detector resolution 
whereas for higher $\Delta m_{41}^2 \geq 0.5$ eV$^2$, the active-sterile neutrino 
mixing sensitivity improves by $\sim$23\% for the detector resolution of 5\% from 
25\% at $\Delta m_{41}^2$ = 1.0 eV$^2$. Here it is to be mentioned that the precision 
on the $\Delta m_{41}^2$ is controlled by the precise measurement of  energy  
(and also $L$) for individual events which depends on the resolution of the detector.
Further studies are carried out considering the detector resolution of $\sigma/E$ = 
20$\%$ (which is the energy resolution of the PS obtained from the 
measurements)~\cite{Mulmule:2018efw}. We have also studied the 
active-sterile neutrino mixing angle, $\sin^22\theta_{14}$ sensitivity by varying the 
detector efficiencies from 20$\%$ to 50$\%$ as shown in Fig~\ref{fig:deteffi}. It is 
observed that the sensitivity on $\sin^22\theta_{14}$ improves with increase of 
detector efficiency for $\Delta m_{41}^2 \leq 4.0$ eV$^2$ and has less impact beyond 
this value. For higher values of $\Delta m_{41}^2$ the oscillation probability washed 
out. From this study, it is concluded that with better detector response, we 
will have better sensitivity in both the active-sterile neutrino mixing parameters 
sin$^22\theta_{14}$ and $\Delta m_{41}^2$. In the subsequent analyses, we have used 
energy resolution as $\sigma/E$ = 20\%/$\sqrt{E}$ and efficiency of 25\% unless 
otherwise stated.

\subsection{REACTOR CORE TO DETECTOR DISTANCE}
The distance between the reactor and the detector is not uniquely defined 
because of the extended reactor core such as DHRUVA reactor. 
Figure~\ref{fig:diffpaths} shows the active-sterile neutrino mixing sensitivity at 
source to detector distances of 7 m, 10 m, and 13 m. These distances correspond the
center to center distance between the reactor core and center of the detector. In 
our calculation  neutrino vertices are generated randomly in the reactor core using 
MC method and assumed a point detector, for a given energy resolution, thermal power 
(100 MW$_{th}$) and, duty cycle of 70\%. The lower limit on source to detector 
distance (7 m) is based on the closest accessible baseline available to place the 
detector. It can be observed that at $\Delta m_{41}^2$ = 1.0 eV$^2$, the 
active-sterile neutrino mixing sensitivity $\sin^22\theta_{14}$ improves by 
$\sim$53$\%$ for the path length of 7 m from 13 m. In addition, one can maximize the 
event statistics and experimental sensitivity by placing the detector close to the 
reactor, however there is a trade-off between distance, other shielding material 
structures surrounding the reactor core and associated reactor background. The green 
dashed-dot line shows the sensitivity on mixing parameters by considering extended 
source as well as detector with centre to centre distance between reactor core and 
detector is 13 m. Both neutrinos production and their interaction point in the 
detector are generated on MC basis. Hence the the closest neutrinos can have a path 
of less than $\sim$11 m and the farthest ones oscillate for more than $\sim$15 m for 
given both the detector and reactor geometries. It is found that with extended 
detector, the sensitivity of the ISMRAN further reduces in the range of  0.3 eV$^2$ 
$ \leq \Delta m_{41}^2 \leq$ 4.0 eV$^2$ as compared to case with extended source 
and point detector placed at distance of 13 m. 

\subsection{REACTOR POWER AND DUTY CYCLE}
The antineutrino flux emitted from the reactor is proportional to its operating 
thermal power. The DHRUVA research reactor~\cite{Agarwal:dhruva} can operate at a
maximum thermal power of 100~MW$_{th}$, where as PFBR power 
reactor~\cite{Chetal:pfbr} can operate at a maximum thermal power of 1250 MW$_{th}$ 
which is an order of magnitude higher than research reactor. 
Figure~\ref{fig:diffpower} shows the comparison of exclusion limits on $\Delta 
m_{41}^2$ - $\sin^22\theta_{14}$ plane at various reactor thermal power of 
60~MW$_{th}$, 100~MW$_{th}$, and 1250 MW$_{th}$ for 1~ton-yr of detector exposure 
time at a distance of 13 m. With the increase in thermal power, there is an increase 
in event statistics hence increase in sensitivity of the experiment at all $\Delta 
m_{41}^2$.

\bef[ht!]
\bc
\includegraphics[width=0.5\textwidth]{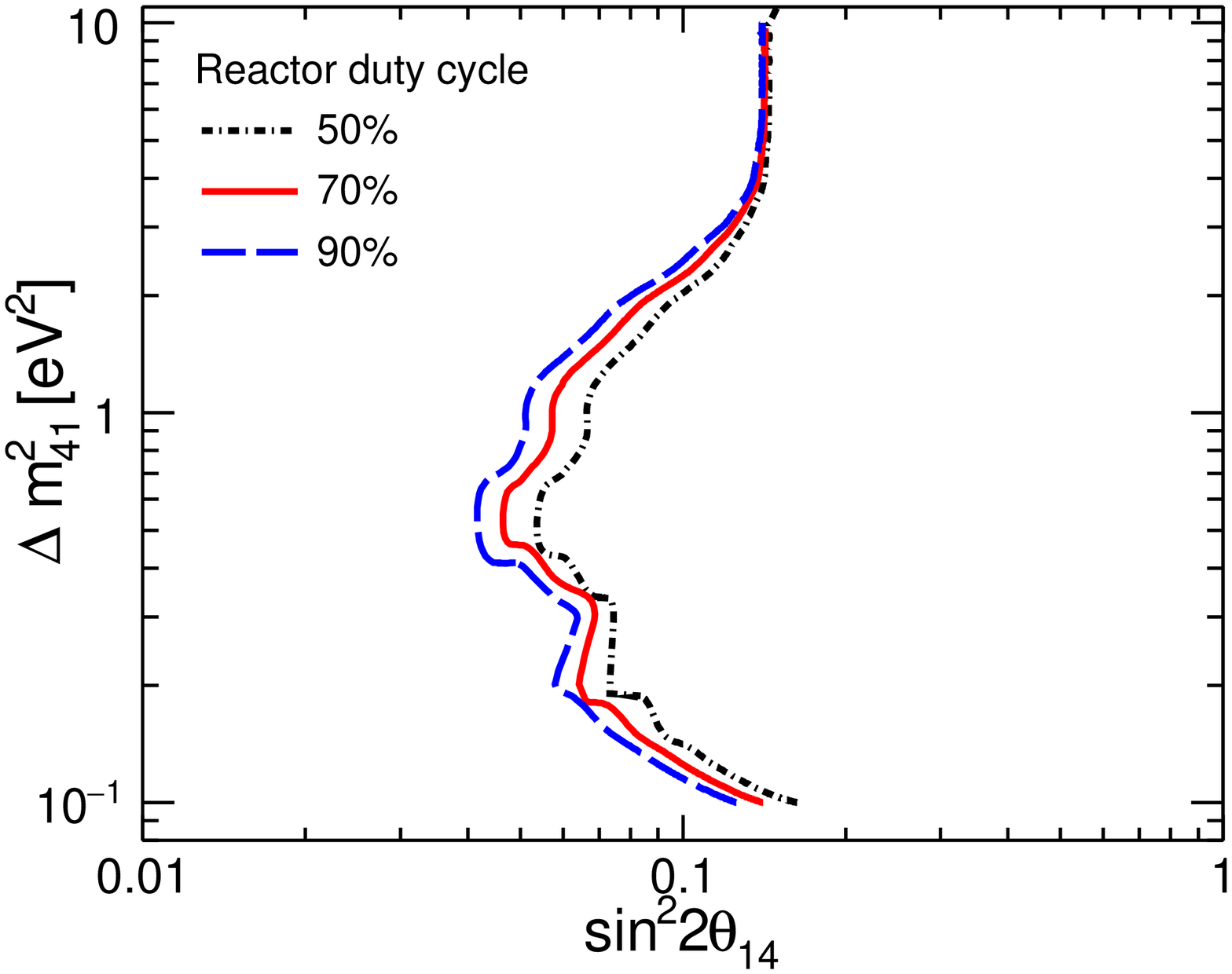}
\caption{The 90\% C.L. exclusion limits in the $\Delta 
m_{41}^2-\sin^22\theta_{14}$ plane, where $\sin^{2}2\theta_{14}= 4U^{2}_{e 
4}(1-U_{e4})^{2}$, expected from 1 ton-yr of the data at different duty cycle of the
reactor.}
\label{fig:dutycycle}
\ec
\eef

\bef[ht!]
\bc
\includegraphics[width=0.5\textwidth]{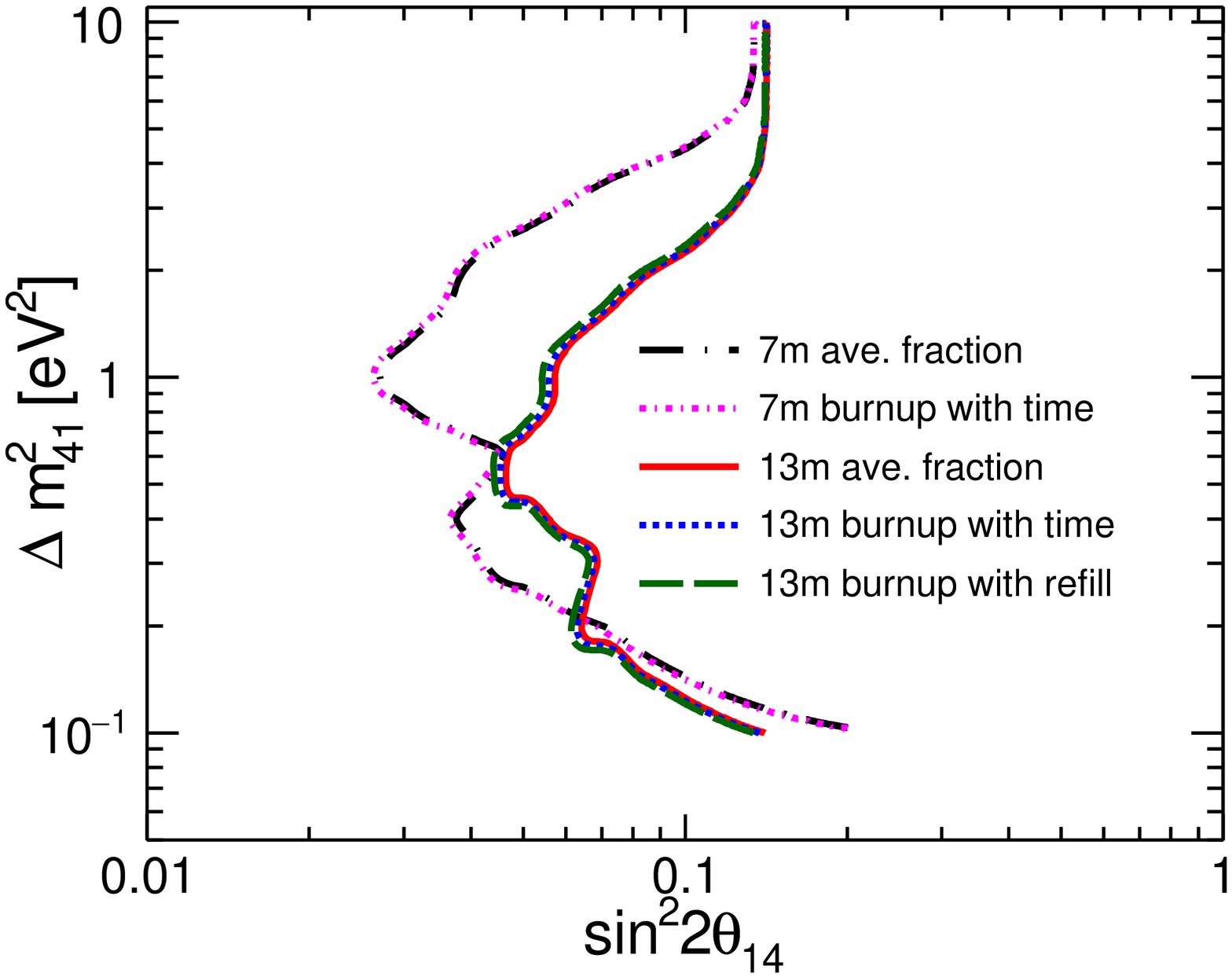}
\caption{The comparison of 90\% C.L. exclusion limits in the $\Delta 
m_{41}^2-\sin^22\theta_{14}$ plane, where $\sin^{2}2\theta_{14}= 4U^{2}_{e 
4}(1-U_{e4})^{2}$, expected from 1 ton-yr of the data at different reactor fuel
evolution.}
\label{fig:burnup}
\ec
\eef
Due to the operation of the nuclear reactors below than its maximum thermal output 
and reactor-off period, the total \antinue~ event statistics gets affected. Hence, 
sensitivity of the sterile neutrino oscillation decreases with lower duty cycle. 
Figure~\ref{fig:dutycycle} shows the active-sterile neutrino mixing sensitivity of 
the detector at reactor duty-cycles of 50\%, 70\% and 90\% for source to detector 
distance of 13 m and, 100 MW$_{th}$ reactor (extended source) thermal power. It is 
observed that the active sterile neutrino mixing sensitivity improves with the duty 
cycle for $\Delta m_{41}^2 <$ 2.0~$eV^2$ and beyond this there is no effect. This is 
due to the averaging out of oscillation probabilities at higher $\Delta m_{41}^2$.

\subsection{REACTOR FUEL}
 Above studies are carried out assuming time averaged constant fission fraction 
contributions of various fuel elements as mentioned in
Table \ref{tab:para} to the thermal power. However, the study has also been carried 
out considering the time evolution of the fissile fraction contribution 
to the reactor thermal power. The time evolution of nuclear 
reactor depends on its fuel cycles which can go from about a month to one or two 
years. In each of the new cycle, a partial or complete fuel is replaced with fresh 
fuel, which is typically composed by enriched $^{235}$U. At the beginning of each 
reactor cycle, the $\antinue$ flux comes mainly from the fission of $^{235}$U, with a 
small contribution of $^{238}$U isotope. The neutron flux produced from the fission 
helps to produce $^{239}$Pu and a small quantity of $^{241}$Pu. Hence, as $^{235}$U 
is consumed with time, then its contribution to the $\antinue$ flux decreases, 
whereas the contributions from $^{239}$Pu and $^{241}$Pu increase. However, the 
dominant contribution comes from the $^{239}$Pu, which is comparable with the 
$^{235}$U towards the end of each cycle~\cite{Giunti:2017yid}. Since our reactor is 
of CANDU type, we have used the burn-up as given in Ref.~\cite{Francis}.
Figure~\ref{fig:burnup} shows the comparison of active-sterile neutrino mixing 
exclusion limits between fixed fission fraction~\cite{Zhan:2008id} and with the 
fission fraction variation due to burn up of the fissile element~\cite{Francis} at 
reactor thermal power 100~MW$_{th}$ for source to detector path lengths of 7 m and 
13 m. It has been observed that the burn up variation of reactor fuel has marginal 
effect on the active-sterile neutrino mixing sensitivity for all considered values 
of $\Delta m_{41}^2$ for this assumed fuel cycle. However, burn-up effect may be 
observed for longer duration of fuel cycle. Also we have shown the effect of fuel 
cycle which is assumed to be 100 days in our case, on the active-sterile neutrino 
mixing and it shows a similar sensitivity with respect to time variation 
reactor burn up.

\bef[ht!]
\bc
\includegraphics[width=0.5\textwidth]{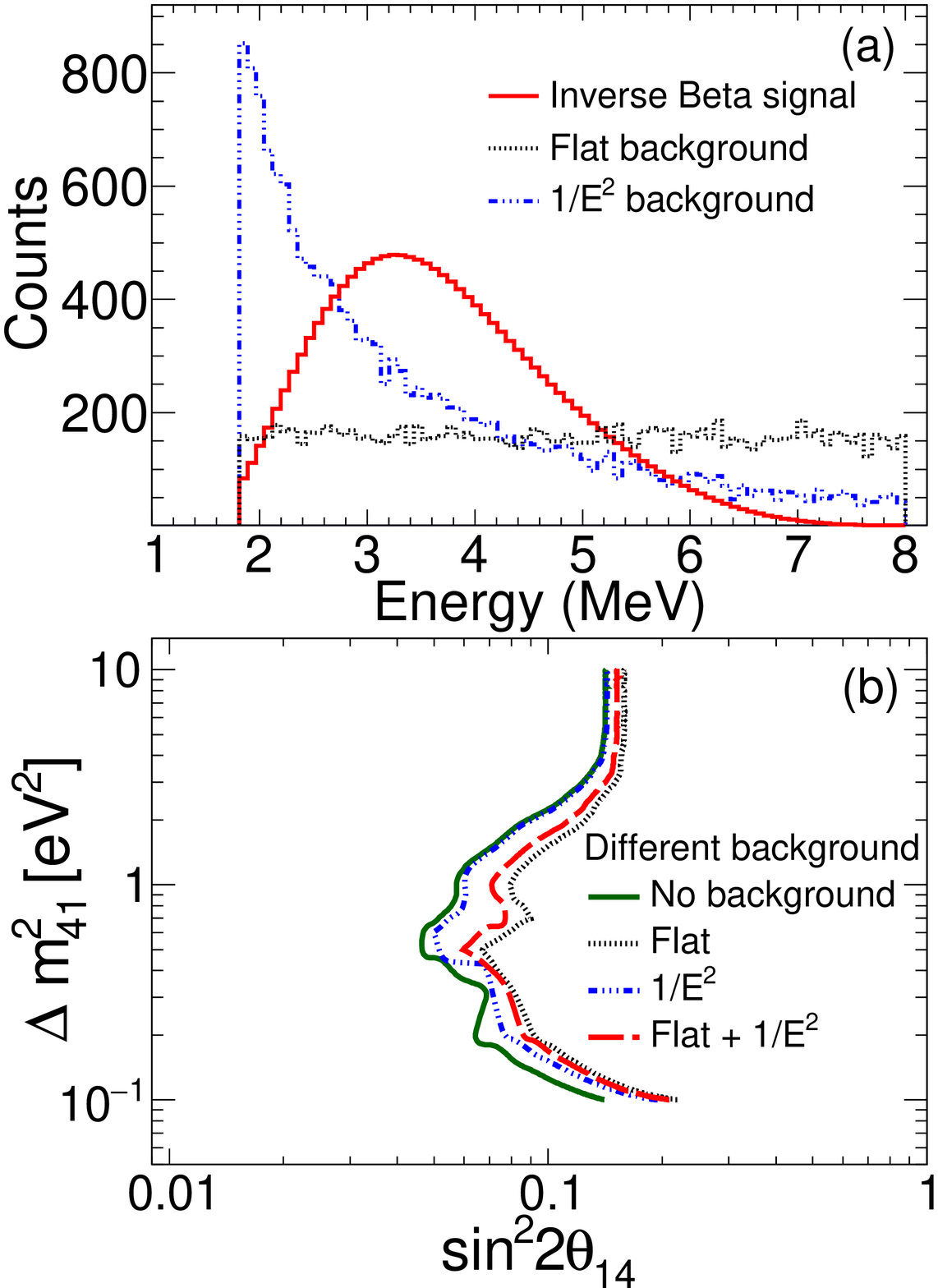}
\caption{IBD like events and different background energy
spectra (in a). The comparison of 90\% C.L. exclusion limits in the
$\Delta m_{41}^2-\sin^22\theta_{14}$ plane, where $\sin^{2}2\theta_{14}= 4U^{2}_{e4}(1-U_{e4})^{2}$,
expected from 1 ton-yr of the data for various background shapes (in b).}
\label{fig:exclbkg}
\ec
\eef

\bef[ht!]
\bc
\includegraphics[width=0.5\textwidth]{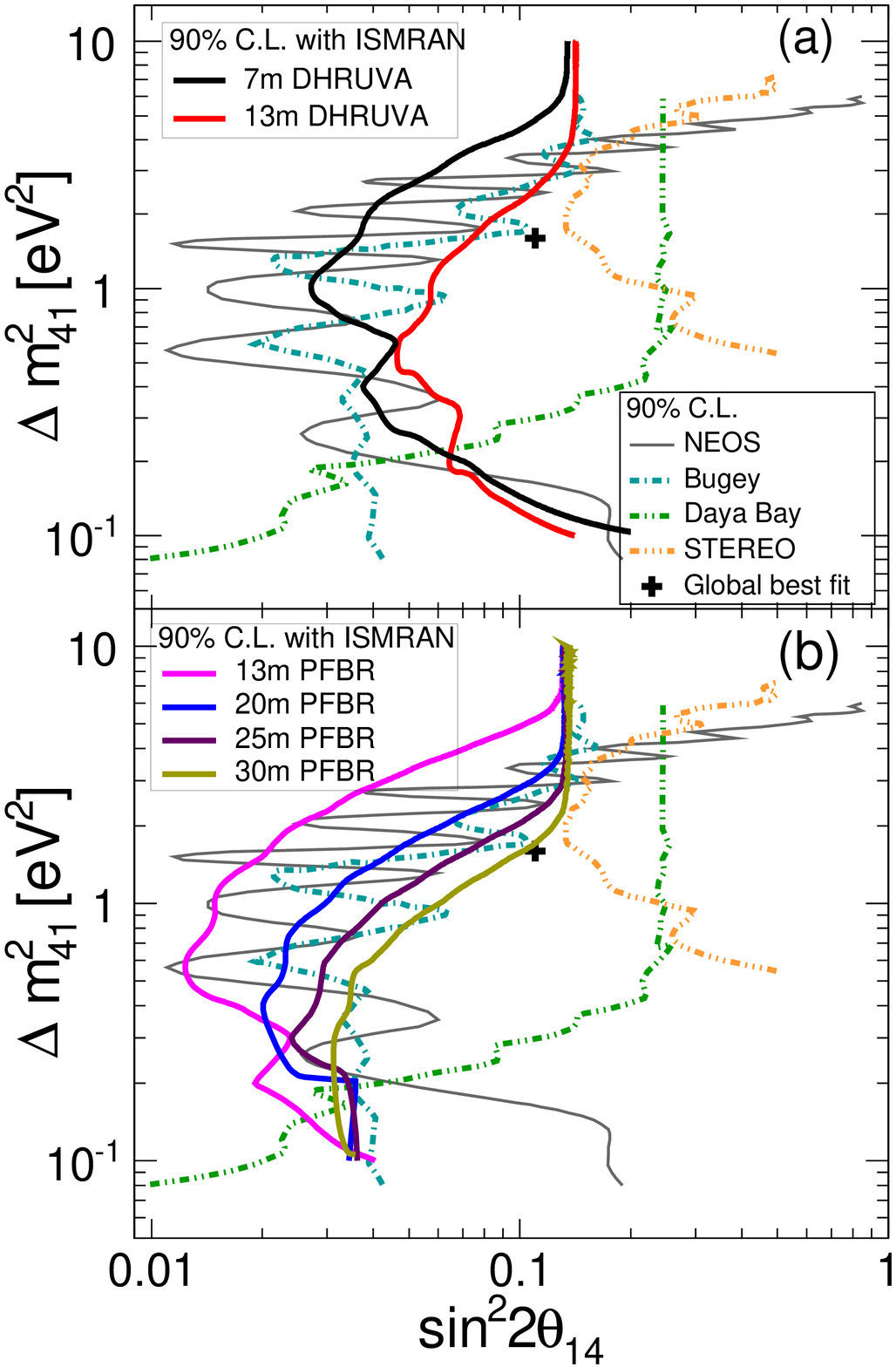}
\caption{The comparison of 90\% C.L. exclusion limits in the $\Delta 
m_{41}^2-\sin^22\theta_{14}$ plane, where $\sin^{2}2\theta_{14}= 4U^{2}_{e 
4}(1-U_{e4})^{2}$ between ISMRAN and other experiments.}
\label{fig:exclucomparison}
\ec
\eef
\subsection{BACKGROUNDS}
The active-sterile neutrino mixing sensitivity has been obtained with
inclusion of backgrounds with an assumption of signal to background ratio
is 1. In the analysis, three different cases of background has been considered such 
as the default 1/$E^2$ shape represents the spectral shape provided by 
accidental backgrounds due to contribution from intrinsic detector radioactivity, a 
flat distribution in antineutrino energy 
due to fast neutron backgrounds~\cite{Heeger:2012tc} and the combination of both 
these backgrounds shown in Fig.~\ref{fig:exclbkg}(a). In this study, an associated 
10$\%$ systematic uncertainty is considered due to these backgrounds. 
Figure~\ref{fig:exclbkg}(b) shows the comparison of ISMRAN detector sensitivity with 
and without inclusion of different backgrounds. It is observed that with the 
contribution of both backgrounds, the active-sterile neutrino mixing angle 
sensitivity is further reduced by $\sim$20$\%$ at $\Delta m^{2}_{41}$ = 1.0 eV$^2$ 
for the case of detector placed at 13~m from the 100 MW$_{th}$ DHRUVA reactor core.
\subsection{COMPARISON TO THE OTHER MEASUREMENTS}
The exclusion limits at 90\% C.L. on the $\sin^22\theta_{14}$ value for each 
$\Delta m_{41}^2$ obtained from the ISMRAN set-up at two different reactors are 
shown in Fig.~\ref{fig:exclucomparison}. It can be noted here
that the analysis has been carried out by smearing both the extended 
source and detector volumes. Neutrinos production in the
reactor core and their interaction in the detector
are generated on MC basis.
 The upper panel shows the results obtained
assuming the detector set-up placed at distance of 7 m and 13 m from
the core of DHRUVA reactor and the lower panel shows sensitivity of
the detector by placing at different distances in the PFBR facility. 
Measurements from other experiments such as, the NEOS~\cite{Ko:2016owz}, the Daya 
Bay~\cite{Adamson:2016jku}, Bugey-3~\cite{Declais:1994su}, 
STEREO~\cite{Almazan:2018wln} and the symbol `$+$' is the present best fit value from 
the global analysis~\cite{Gariazzo:2017fdh} are also shown for comparison at 90$\%$ 
C.L. The results from ISMRAN at a distance of 13 m from DHRUVA reactor core is 
comparable to the NEOS results at lower $\Delta m^{2}_{41}$ $<$ 2 eV$^2$, at higher 
$\Delta m^{2}_{41}$ our results are comparable with the Bugey results and outperform 
the Daya Bay results for $\Delta m_{41}^2 >$ 2 eV$^2$. At a distance of 7 m from the 
reactor core and $\Delta m_{41}^2 >$ 1 eV$^2$, the results from ISMRAN are comparable 
with NEOS and Bugey. The ISMRAN has better sensitivity on the active-sterile 
neutrino mixing with respect to STEREO~\cite{Almazan:2018wln}. The exclusion plot 
from the Daya Bay~\cite{Adamson:2016jku} experiment at lower values of $\Delta 
m^{2}_{41}$ = 0.1 eV$^2$ has better sensitivity compared to ISMRAN and also other 
measurements. However, it is found that the active-sterile neutrino sensitivity of 
ISMRAN improves substantially if the measurement will be carried out at PFBR facility 
as shown in Fig.~\ref{fig:exclucomparison}(b). The ISMRAN results at a distance of 
20 m from the reactor core are comparable to NEOS and Bugey at all values of 
$\Delta m_{41}^2$ and exclusion limits are better for $\Delta m_{41}^2 <$ 2 eV$^2$. 
It is to be noted that NEOS measurements are performed at a distance of $\sim$24 m from the 
reactor core with thermal power of about 3 GW$_{th}$. Although the reactor power of PFBR is 
lower compared to reactor used for NEOS measurements, the results from ISMRAN at PFBR 
can give a better sensitivity as compared to other measurements. This is due to the 
compact core size of the PFBR facility. 
\section{SUMMARY}
\label{sec:summary}
In the near future, results form various SBL experiments using reactor neutrino as 
a source may resolve the uncertainty for the existence of light sterile neutrino 
hypothesis as the possible origin of the RAA and in addition it may clarify the 
origin of the 5 MeV distortion in the $\overline{\nu}_e$ energy spectra. The 
feasibility study on active-sterile neutrino mixing sensitivity is performed 
with the upcoming ISMRAN experimental set-up for an exposure of 1 ton-year employing 
$\overline{\nu}_e$ produced from the extended core of DHRUVA and compact core of PFBR 
reactor facility, India. The study is carried out considering both reactor as well as 
detector related parameters. With varying the source to detector distance of 7m from 
13m at $\Delta m^{2}_{41}$ = 1.0 eV$^2$, the sensitivity on $\sin^{2}2\theta_{14}$ 
improves by twice. It has been observed that, the burn-up variation of the 
reactor fuel elements has very less impact on active-sterile neutrino mixing 
sensitivity. At reactor power of 100 MW$_{th}$ produced from DHRUVA reactor, the 
experimental set up may see the active-sterile neutrino mixing sensitivity if 
$\sin^{2}2\theta_{14} \geq$ 0.064 at $\Delta m^{2}_{41}$ = 1.0 eV$^2$. 
On the otherhand, there is an improvement on the active-sterile 
neutrino mixing parameter $\sin^{2}2\theta_{14}$ to $\sim$ 0.03 for the same 
$\Delta m^{2}_{41}$ by putting the ISMRAN detector set-up at PFBR facility.
We have found the limit on active-sterile neutrino mixing parameters is of the same 
order as that of Bugey. However, in the range of 0.2 eV$^2~< \Delta m^{2}_{41}<$ 3.0 
eV$^2$, the present analysis for DHRUVA reactor predicts the same sensitivity limits 
as that of the results from NEOS. Further, the sensitivity at lower values of $\Delta 
m^{2}_{41} <$ 0.2 eV$^{2}$, we may have better sensitivity compared to NEOS. With 
the ISMRAN set-up, it can be possible to verify the existence of active to sterile 
neutrino oscillation hypothesis as the possible origin of the RAA and, also
to clarify the origin of the bump at 5 MeV in the \antinue~ spectra. 
\\
\section*{ACKNOWLEDGMENTS}
We thank A. K. Mohanty, V. M. Datar and Anushree Ghosh for their helpful 
suggestions and useful discussions. We also thank ISMRAN group members for useful 
discussion.


\end{document}